%% file: paper.tex
\documentclass[a4paper]{jpconf}

\usepackage{graphicx}
\usepackage{amsmath}
\usepackage{amssymb}
\usepackage{subfigure}
\usepackage{epsfig}
\usepackage{wrapfig}

\include{alias}

\begin{document}

\title{Complete collisions approximation to the Kadanoff--Baym equation: a first--principles implementation.}

\author{Davide Sangalli}
\address{CNR-ISM, U.o.s. di Montelibretti, via Salaria Km 29.3, I-00016 Montelibretti, Italy, European Union}
\ead{davide.sangalli@ism.cnr.it}

\author{Andrea Marini}
\address{CNR-ISM, U.o.s. di Montelibretti, via Salaria Km 29.3, I-00016 Montelibretti, Italy, European Union}

\begin{abstract}
We show carriers dynamics on bulk Silicon in the sub pico--second time scale.
The results are obtained from a first--principles implementation of 
the the Kadanoff--Baym equations within the
generalized Baym--Kadanoff ansatz and the complete collision approximation.
The resulting scattering term is similar to the scattering described within
the semi--classical Boltzmann equation~\cite{Haug2}.
\end{abstract}

\section{Introduction}

The description of carriers dynamics in semi--conductors
is crucial for a correct modelization of electronic devices. Silicon ($Si$) in
particular is the fundamental building block of semi--conductors physics and
microelectronics industry~\cite{Omara1}.

The reference equation used in the literature is the semi--classical Boltzmann equation (sc-BE)
which describes the scattering of electrons and holes with
lattice vibrations, electron--phonons ({e--p}) scattering, with other carriers, electron--electron ({e--e})
scattering and possibly with defects. Its implementation is usually based on models,
where the characteristic of the system are taken into account with ad--hoc parameters.
In the present work we want instead to describe such scattering events within a 
first--principles approach.

Also the sc--BE assumes a semi--classical regime, as opposite to the quantum--kinetic regime
which should, in principle, describe the electrons propagation. 
However the miniaturization of $Si$--based devices to the nano--scale regime and the never ending
search for faster devices call for a deep understanding of the fundamental quantum--mechanical process
that governs the carriers dynamics~\cite{Haug1,Sundaram1}.
Moreover the recent development of ultra--short laser pulses has triggered the necessity to
understand the quantum kinetics of electrons and holes on the sub pico--second ($ps$) time--scale.

The non--equilibrium Green's functions (NEGF) approach offers a quantum--kinetic equation,
the Kadanoff--Baym equation (KBE), which describes the quantum dynamics through the
many--body Green's function $G^<(t,t')$. The solution of the full
equation however is not feasible within a first--principles approach, and also for simple models,
an approximate version is usually adopted. The most common approximation is the
generalized Baym--Kadanoff ansatz (GBKA), which reduces the KBE to a closed equation for the
time--diagonal Green's function, i.e. the time--dependent density matrix of the system.
If also the complete--collision approximation (CCA) is employed the scattering processes
are described at the same level of the sc--BE~\cite{Haug2}, however with a smeared 
energy conservation due to possible violations on the ultra--short 
time regime. We will refer to this equation as the CCA--KBE.

The CCA--KBE has been recently implemented in the yambo~\cite{yambo_code} code.
Projected onto the Kohn--Sham (KS) wave--functions, it 
describes the evolution of a parameter free $G^<_{ij}(t,t)$,
which can be used to reconstruct the electronic occupations: ${f_i=-i G^<_{ii}(t,t)}$.
Here we used the generalized index $i=\{n\mathbf{k}\sigma\}$. 
Although the scattering terms still have a semi--classical form,
the CCA--KBE offers a suitable starting point for future development towards a
real quantum kinetics equation. It also has other
advantages. For example it can correctly model the interaction with
ultra--short laser pulses. Indeed the first--principles CCA--KBE has been shown to
reproduce the propagation of the polarization~\cite{Attaccalite1},
${P=\sum_{i\neq j}G^<_{ij}(t,t)}$ induced by a laser pulse.

In the present work we explore the performance of its implementation on bulk Silicon
and in particular we will focus our attention of the scattering term $S_i(t)$.
The details and the approximations needed for a first--principles implementation of this term
have been discussed in Ref.~\cite{Marini1}. 

\section{The theoretical method}
The theoretical approach used is described in Refs.~\cite{Attaccalite1,Marini1}.
Here we report the main equations. The CCA--KBE equation reads
\begin{equation} \label{eq:G_lesser_EOM}
i\hbar\partial_t G^<_{ij}(t) - \left[ H_{MB} + H_{RT}(t) , G^<(t)\right]_{ij} = S_{ij}(t)
\end{equation}
with
\begin{subequations}
\begin{eqnarray}
&&H_{DFT}[\rho^{eq}]   = H_{0}+V_H[\rho^{eq}]+V_{xc}[\rho^{eq}] \\
&&H_{MB}[G^{<,eq}]     = H_{DFT}[\rho^{eq}]+(\Sigma[G^{<,eq}]-V_{xc}[\rho^{eq}]) \\
&&H_{RT}[G^{<}(t)](t)  = V_H[\delta\rho(t)] + \left( \Sigma[\delta G^<(t)],G^{<,eq}]\right)+U^{ext}(t) \\
&&S(t)                 = \{\Sigma_C^>,G^<\}+\{\Sigma_C^<,G^>\}
\end{eqnarray}
\end{subequations}
$H_{DFT}$ is the standard KS Hamiltonian. ${H_{MB}}$ describes the quasi--particle (QP) corrections. Finally
the ${H_{RT}}$ Hamiltonian instead includes the external field
and the variation in the Hartree potential and in the self--energy~\cite{Attaccalite1}.
In the present work we replace its role with an ad--hoc initial guess for the electrons and
holes populations.

\begin{figure}[th]
\begin{tabular}{llll}
\includegraphics[natwidth=10px,natheight=10px,width=0.22\textwidth]{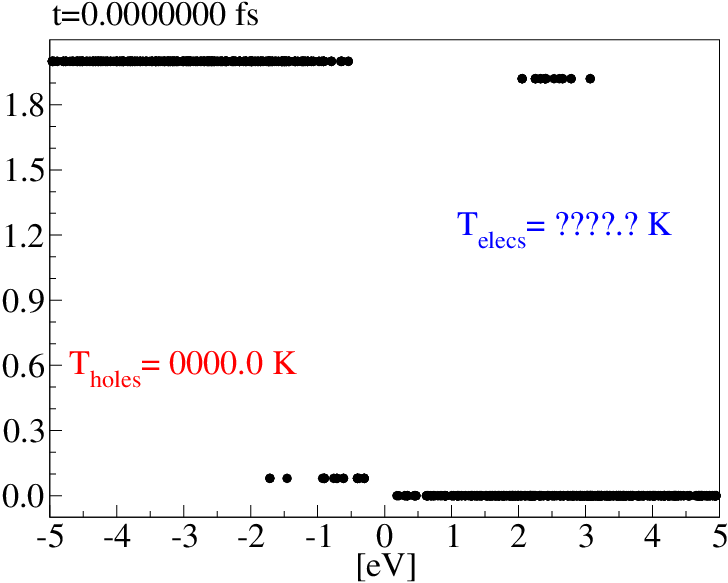}&
\includegraphics[natwidth=10px,natheight=10px,width=0.22\textwidth]{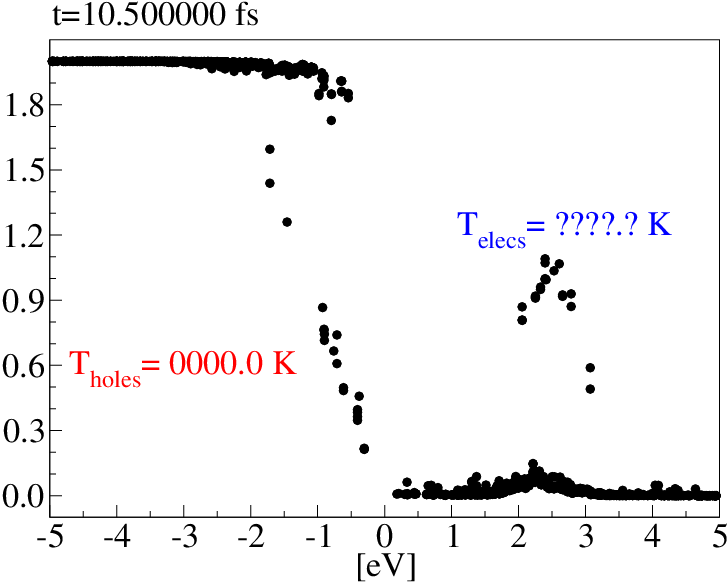}&
\includegraphics[natwidth=10px,natheight=10px,width=0.22\textwidth]{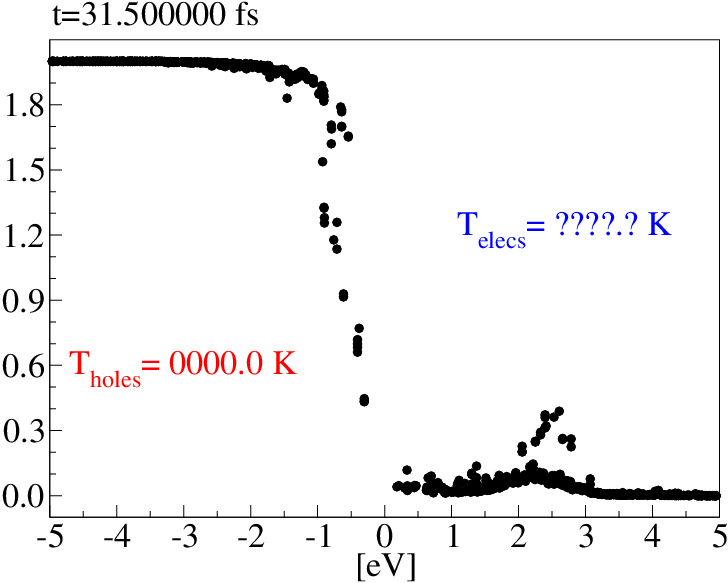}&
\includegraphics[natwidth=10px,natheight=10px,width=0.22\textwidth]{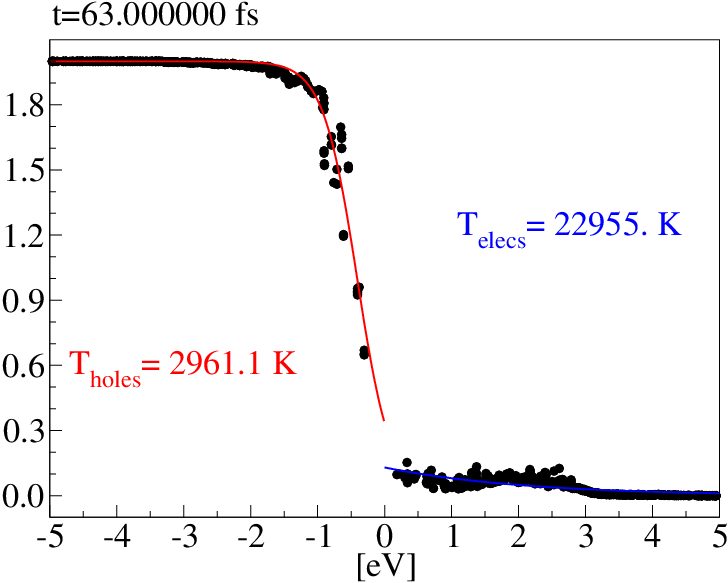}\\
\includegraphics[natwidth=10px,natheight=10px,width=0.22\textwidth]{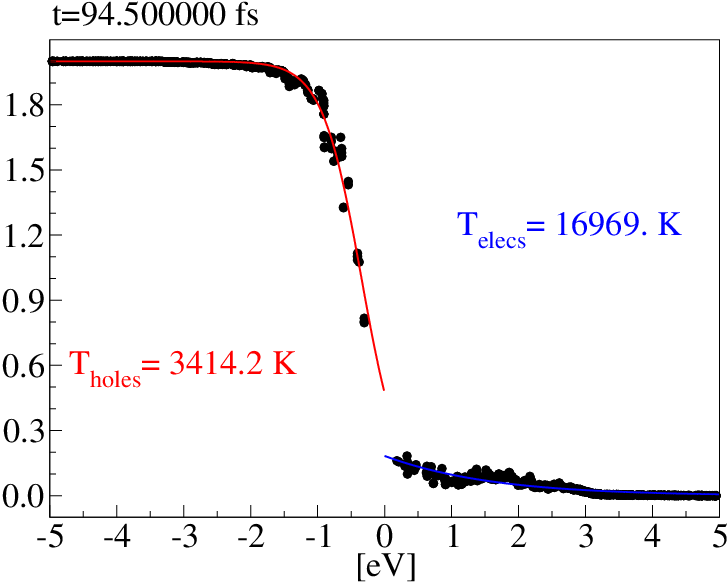}&
\includegraphics[natwidth=10px,natheight=10px,width=0.22\textwidth]{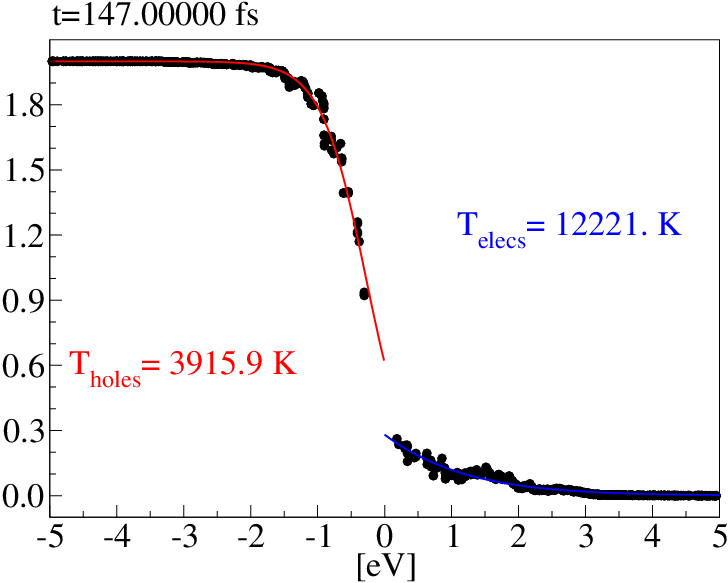}&
\includegraphics[natwidth=10px,natheight=10px,width=0.22\textwidth]{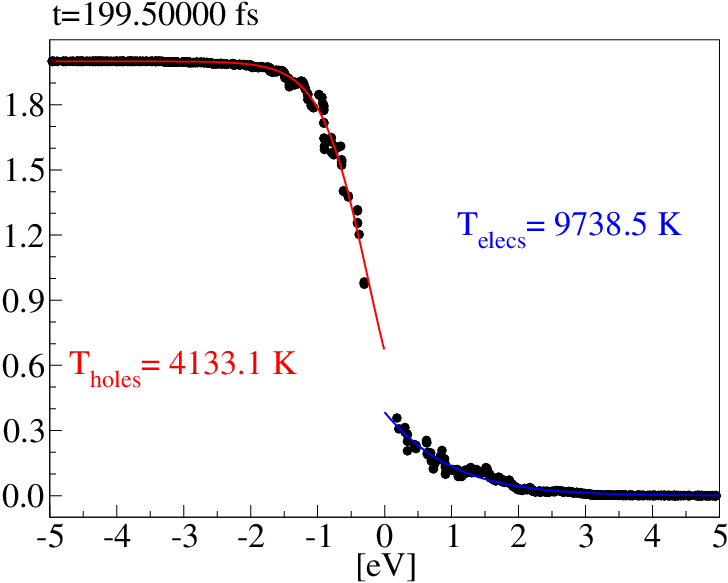}&
\includegraphics[natwidth=10px,natheight=10px,width=0.22\textwidth]{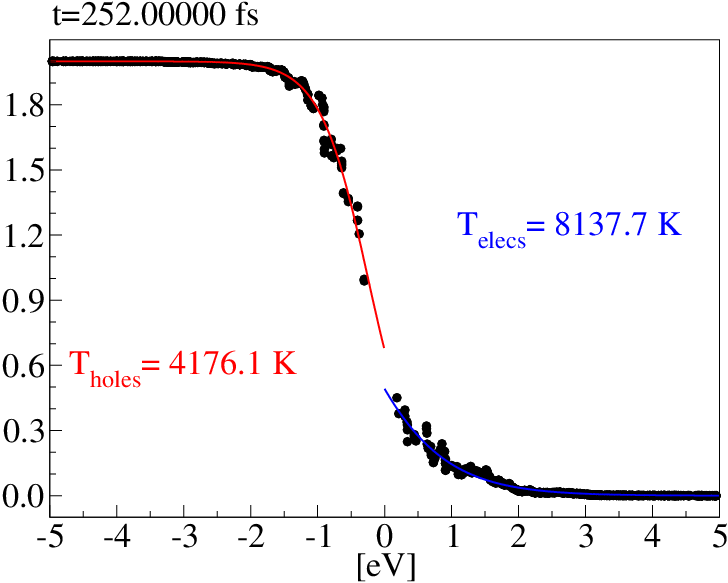}\\
\includegraphics[natwidth=10px,natheight=10px,width=0.22\textwidth]{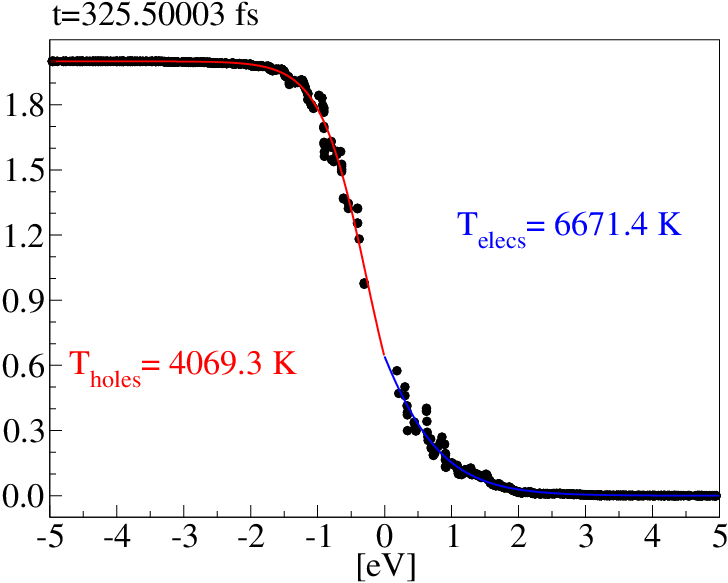}&
\includegraphics[natwidth=10px,natheight=10px,width=0.22\textwidth]{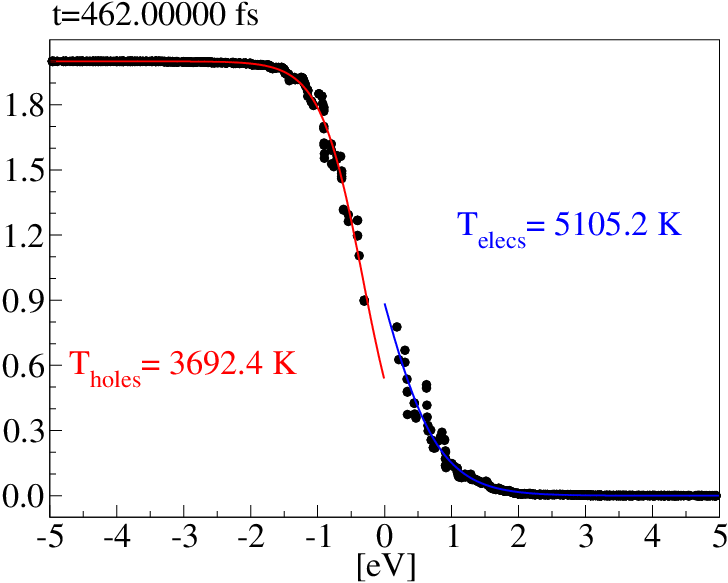}&
\includegraphics[natwidth=10px,natheight=10px,width=0.22\textwidth]{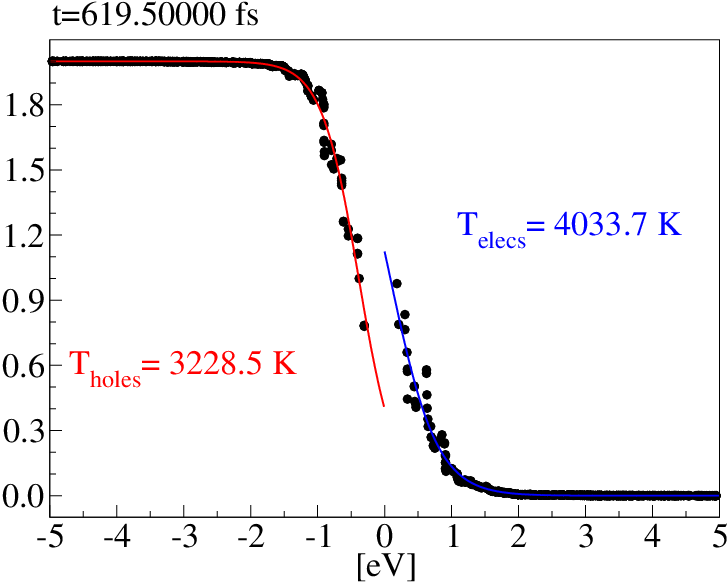}&
\includegraphics[natwidth=10px,natheight=10px,width=0.22\textwidth]{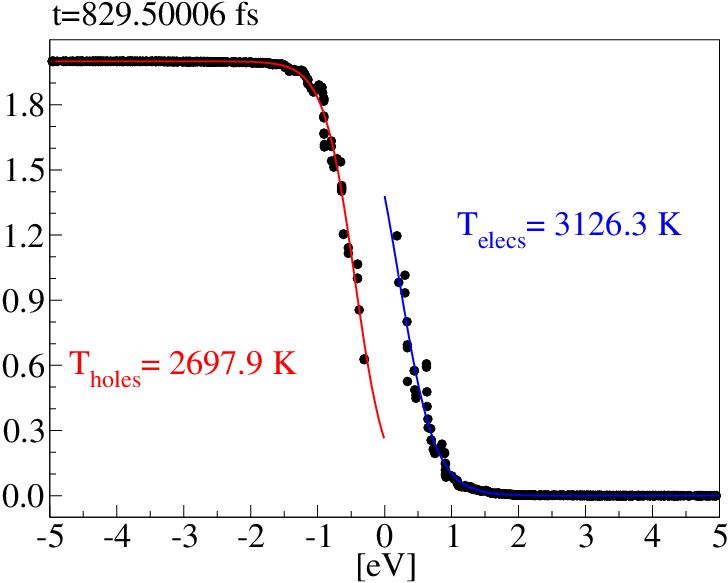}
\end{tabular}
\caption{Carriers occupations distribution as a function of time}
\label{fig:carr_occ_high_pump}
\end{figure}

The term $S(t)$ describes the carriers scattering~\cite{Marini1}.
Since the electronic occupations are the diagonal terms of the $G^<{ij}$,
we focus our attention on the diagonal of eq.~\ref{eq:G_lesser_EOM}.
The latter has only contributions from the $S(t)$ term, since we replace the role of $H_{RT}$ with
an initial guess and $H_{MB}$ gives no contribution on the diagonal.
Indeed the CCA--KBE, together with a QP approximation for the retarded and advanced Green's function~\cite{Marini1},
has the form
\begin{equation} \label{eq:f_t_EOM}
\left.\partial_t f_{i}\(t\)\right|_{relax}=-\gc^{(e)}_{i}(t) f^{(e)}_{i}(t)+\gc^{(h)}_{i}(t) f^{(h)}_{i}(t),
\end{equation}
with ${f^{(e)}_{i}=f_{i}}$ and ${f^{(h)}_{i}=1-f_{i}}$. The electron and hole lifetimes, 
${\gc^{\(e\)}}$ and ${\gc^{\(h\)}}$, include both ({e--p}) and ({e--e})
contribution: ${\gc^{\(e/h\)}=\left.\gc^{\(e/h\)}\right|_{e-e}+\left.\gc^{\(e/h\)}\right|_{e-p}}$.
 ${\gc^{\(e/h\)}=\left.\gc^{\(e/h\)}\right|_{e-e}+\left.\gc^{\(e/h\)}\right|_{e-p}}$.
In the {e--e} case we have that:
\begin{equation} \label{eq:ee_scatt}
\left.\gamma_{i}^{(e/h)}\right|_{e-e}\propto \sum_{j k l} \left| W_{ij\rightarrow kl}\right|^2 P\(\Delta_{ij}-\Delta_{kl}\)\\ f^{\(h/e\)}_j\(t\)  f^{\(h/e\)}_k\(t\)  f^{\(e/h\)}_l\(t\),
\end{equation}
while in the {e--p} case
\begin{equation} \label{eq:ep_scatt}
\left.\gamma_{i}^{(e/h)}\right|_{e-p}\propto \sum_{j\gl}\sum_{I=\pm} N^{\gl}_I\left| g^{\gl}_{i\rightarrow j}\right|^2 P\(\go_{\gl}\pm I \Delta_{ij}\) f^{\(h/e\)}_j\(t\).
\end{equation}
In Eqs.~\ref{eq:ee_scatt}-\ref{eq:ep_scatt} $\Delta_{ij}=\gee_i -\gee_j$ and $\gl$ represents a generic
phonon mode with momentum $\qq$ and branch $\eta$. $N^{\gl}_{+}=\(1+N_\gl\(T\)/2\)$ 
and $N^{\gl}_{-}=N_\gl\(T\)/2$, with $N_\gl\(T\)$ the Bose distribution function at energy $\go_\gl$ and
temperature $T$. In the present work we assume that the phonons distribution does not change
during the dynamics. $W$ is the statically screened component of the Couloumb interaction whose frequency
dependence is embodied in the $P$ functions. These represent a smeared  energy conservation condition. 
Finally, the $g$ are the matrix elements of the screened ionic potential derivative, calculated
within DFPT~\cite{Gonze1995, Baroni2001}.

\section{Results on bulk Silicon}

In the present work we used the local--density approximation (LDA) to describe
bulk Silicon with norm--conserving pseudo--potentials with an energy cet--off of $20\ Ha$.
For the Brilloine zone we use a uniform $4x4x4$
sampling and we consider the first 7 bands, i.e. 4 occupied bands and 3 empty bands. 
The QP corrections are then described with a scissor operator of $0.8\ eV$,
which is known to be a very good approximation for buk $Si$. Then we compute the phonons
and the {e--p} matrix elements with a DFPT run using the same parameters.
The CCA--KBE is then propagated in the KS wave--functions space,
but on an auxiliary $15x15x15$ k--grid. This bigger grid is required
to ensure a proper description of the {e--p} scattering. The KS eigen--values are computed
on this finer grid, while all other quantities
are interpolated from the $4x4x4$ grid with a constant nearest--neighbor interpolation.
The phonons are assumed to remain in equilibrium according to a Bose distribution of $300\ K$.
The time step for the propagation is $dt=10\ as$, while we use a second--order Runge--Kutta integration scheme.
The conservation of the number of carriers during the dynamics is guaranteed by the detailed balance,
as discussed in Ref.~\cite{Marini1}. 

We consider a starting point which mimics a high number of
electrons excited from a laser pulse centered at the optical band--gap of $Si$. Thus we create near the $\Gamma$
point, a packet of holes in valence and a packet of electrons in conduction.
We use here a very high density of carriers,
which would physically lead to a drastic change to the characteristics of the material.
The choice is done to test the implementation in a high--carriers regime.
Indeed here we are interested in exploring two selected physical effects,
i.e. the {e--e} scattering and the {e--p} scattering, in this regime. 
In a lower density regime the effect of the {e--e} scattering would become
smaller and it would be very difficult to test the reliability of the implementation
and eventual errors.

In doing this we will neglect some physical effects, such as the softening of the phonon frequencies,
the change in the band structure due to the extra carriers, the deviation of the phonons distribution
from a Bose function and the sub--sequent heating of the lattice.
All these effects becomes negligible in a lower density regime whereas instead the weaker
{e--e} scattering would still have a role.

In Fig.~\ref{fig:carr_occ_high_pump} we see different time--snapshots of the carriers distribution
as a function of the energy. We see that the electrons and the holes, which at the initial time are
in a highly non--equilibrium situation,
relax towards two Fermi distributions with changing temperatures.
In the holes channel, carriers are closer to their minimum (i.e. the valence band maximum, VBM) and they reach
a lower temperature Fermi distribution. In the electrons channel, instead, we can appreciate the different role played by the
{e--e} scattering and the {e--p} scattering. Phonons have, on average, an energy of few $meV$ and can scatter electrons
only to levels energetically nearby, the result is a packet of electrons which keeps approximately the initial spread
in energy and slowly moves towards the conduction band minimun (CBM). This is poorly visible because meanwhile
the {e--e} channel is able to directly scatter electrons to the CBM.

\begin{wrapfigure}{l}{0.5\textwidth}
\vspace{-20pt}
\begin{center}
\includegraphics[width=18pc]{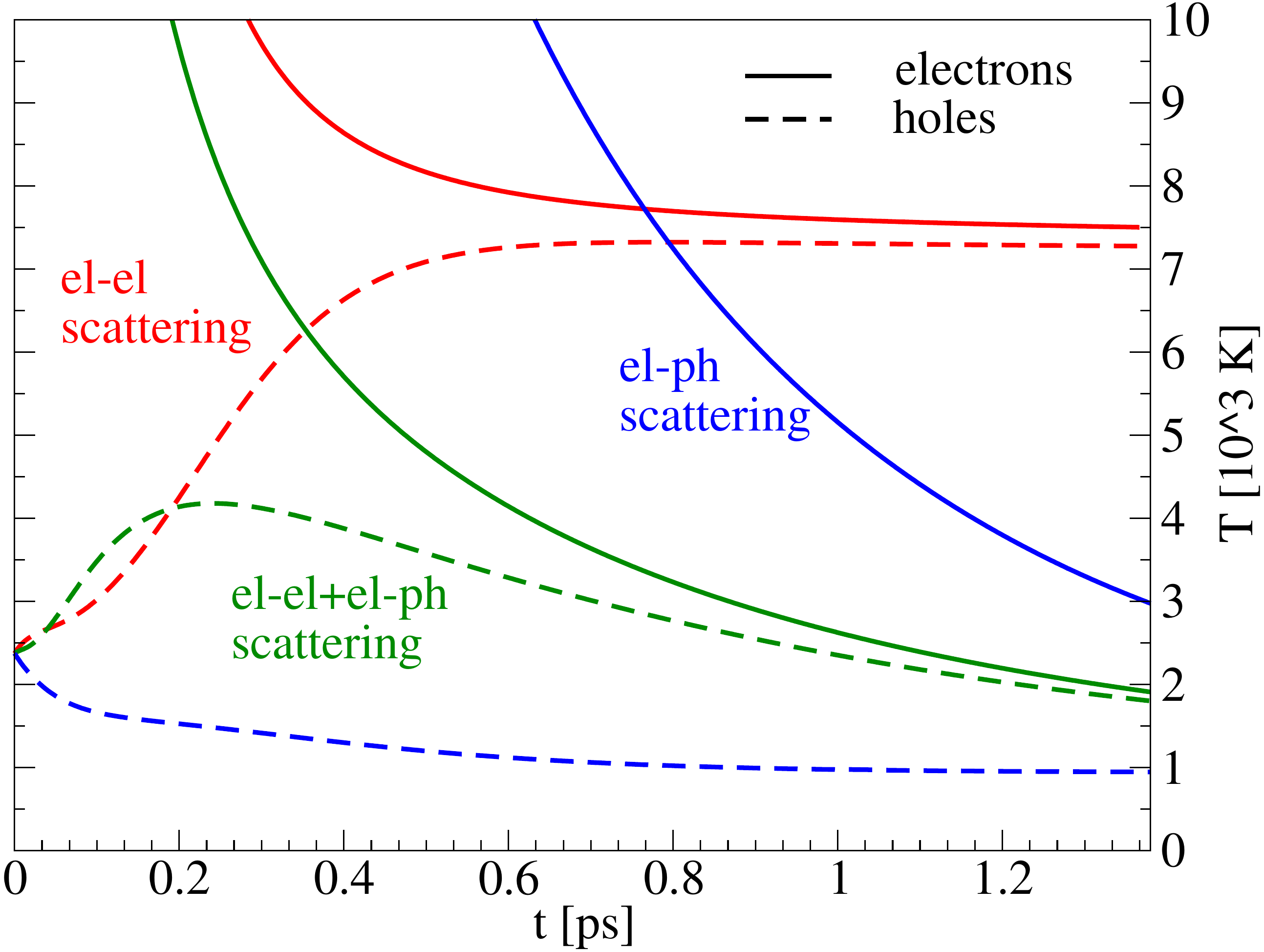}
\end{center}
\vspace{-20pt}
\caption{\label{fig:temp_high_pump}Overview of the electrons and holes temperatures for different scattering mechanisms}
\vspace{-10pt}
\end{wrapfigure}
The resulting temperatures, obtained as the fit of the two Fermi distributions, are reported in Fig.~\ref{fig:temp_high_pump}.
Here, together with the electron and hole temperature for the simulation with both {e--e} and {e--p} scattering on
we report also the fit in case only one of the two relaxation channels is open. The {e--e} scattering (red lines) makes the electrons and holes 
interact and thus increase the holes temperature lowering the electrons temperature. However once the two 
sub--systems reach a thermal equilibrium nothing more can happen. Instead the \mbox{e--p} channel (blue lines) makes the carriers exchange
energy with the lattice. Thus both electrons and holes decrease their temperature. However since phonons do not have 
enough energy to cross the electronic gap, the two processes are independent. Finally, when both scattering channels are open
the resulting effect is a combination of the two. In this high carriers density regime, the {e--e} scattering
dominates the dynamics in the initial part of the dynamics while, once electrons and holes reach a similar temperature,
the {e--p} channel becomes the dominant channel.

The temperatures reached by electrons and holes depends mainly on the initial 
guess (or in case a laser pulse were used on its characteristics) and on the Silicon band structure. They are instead
almost independent on the total number of carriers created. Indeed lowering the carriers density the main change
is that the {e--e} scattering become less effective while all occupations numbers are rescaled. In the limit of
very low carriers concentration the evolution of the carriers temperature becomes very similar to the
case with only the {e--p} channel active (data not shown).

\begin{figure}[t]
\begin{tabular}{lllll}
\includegraphics[natwidth=10px,natheight=10px,width=0.18\textwidth]{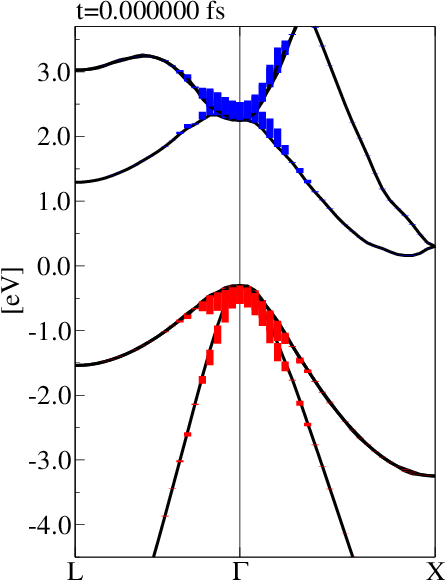}&
\includegraphics[natwidth=10px,natheight=10px,width=0.18\textwidth]{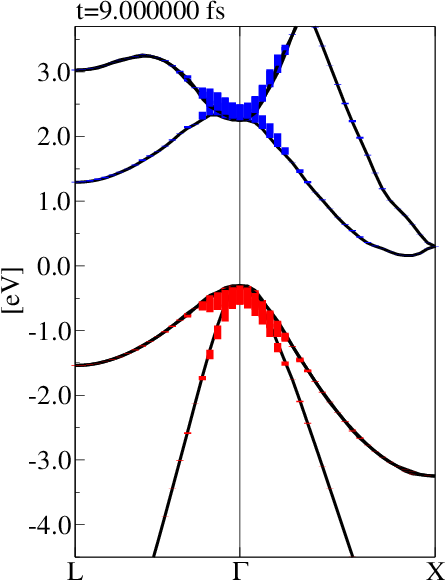}&
\includegraphics[natwidth=10px,natheight=10px,width=0.18\textwidth]{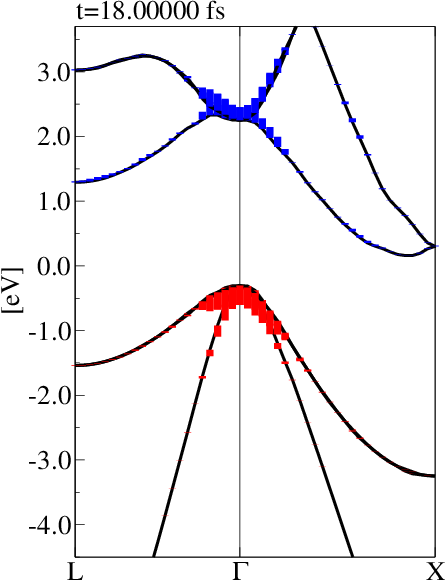}&
\includegraphics[natwidth=10px,natheight=10px,width=0.18\textwidth]{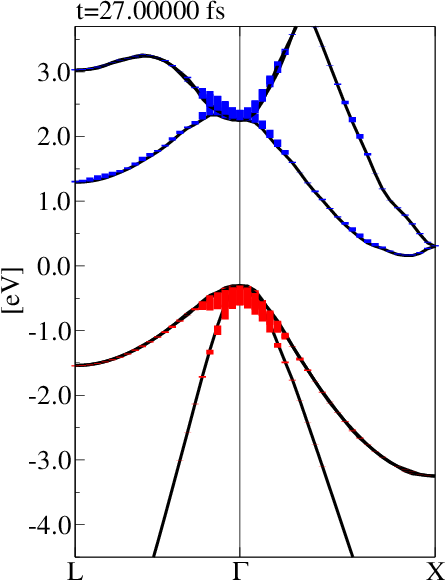}&
\includegraphics[natwidth=10px,natheight=10px,width=0.18\textwidth]{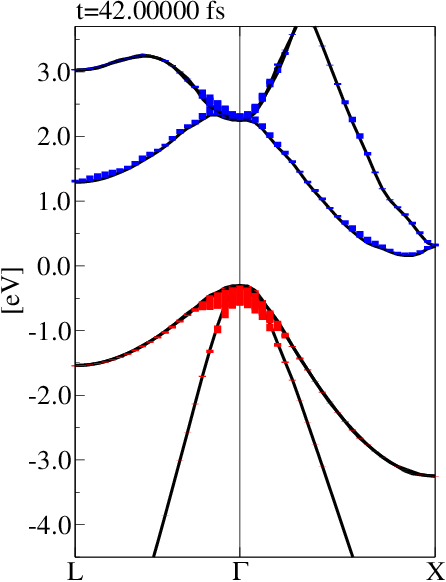}\\
\includegraphics[natwidth=10px,natheight=10px,width=0.18\textwidth]{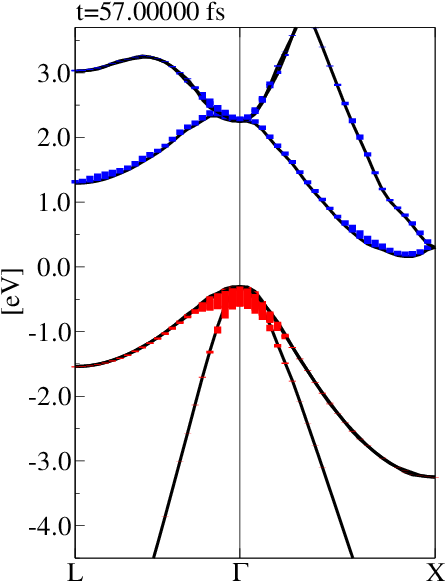}&
\includegraphics[natwidth=10px,natheight=10px,width=0.18\textwidth]{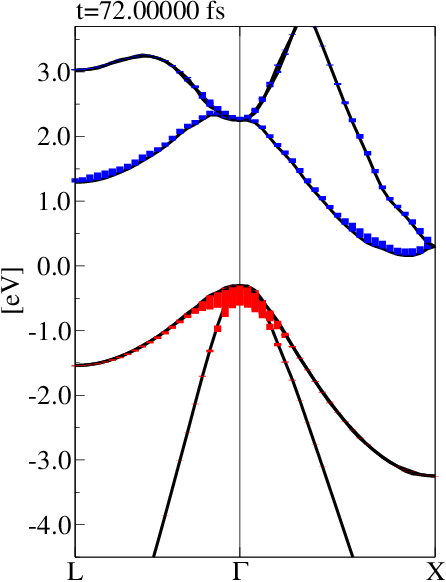}&
\includegraphics[natwidth=10px,natheight=10px,width=0.18\textwidth]{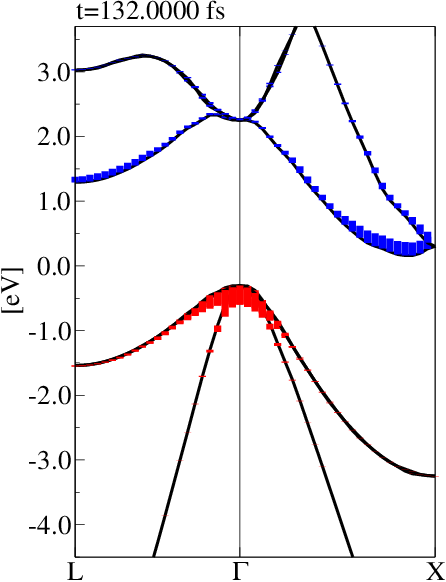}&
\includegraphics[natwidth=10px,natheight=10px,width=0.18\textwidth]{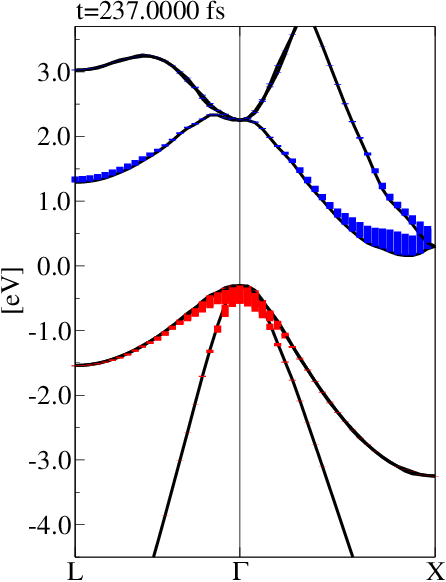}&
\includegraphics[natwidth=10px,natheight=10px,width=0.18\textwidth]{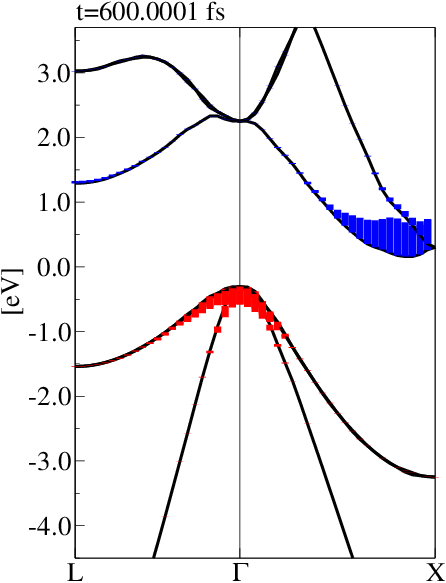}
\end{tabular}
\caption{Carriers occupations on the band structure as a function of time}
\label{fig:carr_bands_high_pump}
\end{figure}

Finally in Fig.~\ref{fig:carr_bands_high_pump} the carriers dynamics is represented on the whole band structure.
Here we see from another perspective the difference between the two channels. Holes remain basically close the VBM,
just cooling down, while electrons move along the band structure going from $\Gamma$ to the minimum, nearby $X$.

In conclusion we have presented a first--principles implementation of carriers dynamics in bulk Silicon.
The most critical aspect of the present simulations is the use of a double grid technique for the sampling of the
Brilloine zone to guarantee a correct description of the energy transfered from electrons to phonons.
The global picture reveal a three systems dynamics: the electrons in the conduction band, the holes in the valence band
and the lattice. The three systems exchange energy, electrons an holes vie the {e--e} scattering, while carriers and
phonons via the {e--p} scattering. Both electrons and holes reach an internal meta--stable state,
i.e. they are distributed according to a Fermi distribution, very quickly (approx 0.1 $ps$).
However the complete equilibration, either in between electrons and holes (less then 1 $ps$)
or with the lattice (not yet reached after 1.5 $ps$), takes longer time. 
This is, to the best of our knowledge, the first fully ab--initio and parameters free
description of carriers dynamics in Silicon.

\section*{References}
\bibliography{paper}

\end{document}

%% file: alias.tex
\renewcommand{\(}{\left(}
\renewcommand{\)}{\right)}

\newcommand{\be}{\begin{equation}}
\newcommand{\ee}{\end{equation}}
\newcommand{\bea}{\begin{eqnarray}}
\newcommand{\eea}{\end{eqnarray}}

\def\qq		{{\bf q}}

\def\qq		{{\bf q}}

\def\gc         {\gamma}

\def\gee        {\epsilon}
\def\gl         {\lambda}

\def\go         {\omega}

\renewcommand{\(}{\left(}
\renewcommand{\)}{\right)}

\def\dk         {\frac{d\,{\bf k}}{\(2 \pi\)^3}}
\def\dk1        {\frac{d\,{\bf k}_1}{\(2 \pi\)^3}}

\def\dk         {\frac{d\,{\bf k}}{\(2 \pi\)^3}}

\def\bverb      {\renewcommand{\baselinestretch}{1}\begin{verbatim}}
\def\everb  	{\end{verbatim}\renewcommand{\baselinestretch}{1.3}}